
\input epsf                                                               %
\input harvmac
\def\Title#1#2{\rightline{#1}\ifx\answ\bigans\nopagenumbers\pageno0\vskip1in
\else\pageno1\vskip.8in\fi \centerline{\titlefont #2}\vskip .5in}

%
%
\ifx\epsfbox\UnDeFiNeD\message{(NO epsf.tex, FIGURES WILL BE IGNORED)}
\def\figin#1{\vskip2in}
\else\message{(FIGURES WILL BE INCLUDED)}\def\figin#1{#1}
\fi
\def\Fig#1{Fig.~\the\figno\xdef#1{Fig.~\the\figno}\global\advance\figno
 by1}
%
%
%
%
\def\ifig#1#2#3#4{
\goodbreak\midinsert
\figin{\centerline{\epsfysize=#4truein\epsfbox{#3}}}
\narrower\narrower\noindent{\footnotefont
{\bf #1:}  #2\par}
\endinsert
}
%
%
\def\calo{{\cal O}}
\def\otp{{1\over2\pi}}
\def\hf{{1\over2}}
\font\ticp=cmcsc10
\def\sq{{\vbox {\hrule height 0.6pt\hbox{\vrule width 0.6pt\hskip 3pt
   \vbox{\vskip 6pt}\hskip 3pt \vrule width 0.6pt}\hrule height 0.6pt}}}
\def\ajou#1&#2(#3){\ \sl#1\bf#2\rm(19#3)}
\def\frac#1#2{{#1 \over #2}}

\def\eg{{\it e.g.}}

\def\p+{{\partial_+}}

\def\ltwid{{\mathrel{\raise.3ex\hbox{$<$\kern-.75em\lower1ex\hbox{$\sim$}}}}}
\def\gtwid{{\mathrel{\raise.3ex\hbox{$>$\kern-.75em\lower1ex\hbox{$\sim$}}}}}
\def\Ssl{{\,\raise.15ex\hbox{/}\mkern-10.5mu S}}
\def\sq{{\vbox {\hrule height 0.6pt\hbox{\vrule width 0.6pt\hskip 3pt
   \vbox{\vskip 6pt}\hskip 3pt \vrule width 0.6pt}\hrule height
0.6pt}}}
\def\ldv{{linear\ dilaton\ vacuum}}
%
%
\lref\Lowe{D. Lowe, ``Semiclassical approach to black hole
evaporation,''\ajou Phys. Rev. &D47 (93) 2446, hep-th/9209008.}
\lref\Jaco{T. Jacobson, ``Black hole radiation in the presence of a short
distance cutoff,'' University of Maryland/ITP preprint
UMDGR93-32=NSF-ITP-93-26.}
\lref\QTDG{S.B. Giddings and A. Strominger, ``Quantum theories of dilaton
gravity,''\ajou Phys. Rev. &D47 (93) 2454, hep-th/9207034.}
\lref\CBHR{S.B. Giddings, ``Constraints on black hole remnants,'' UCSB
preprint UCSBTH-93-08, hep-th/9304027.}
\lref\GiNe{S.B. Giddings and W.M. Nelson, ``Quantum emission from
two-dimensional black holes,''\ajou Phys. Rev. &D46 (92) 2486,
hep-th/9204072.}
\lref\DXBH{S.B. Giddings and A. Strominger, ``Dynamics of Extremal Black
Holes,''\ajou Phys. Rev. &D46 (92) 627, hep-th/9202004.}
\lref\CGHS{C.G. Callan, S.B. Giddings, J.A. Harvey, and A. Strominger,
``Evanescent black holes,"\ajou Phys. Rev. &D45 (92) R1005, hep-th/9111056.}
\lref\BGHS{B. Birnir, S.B. Giddings, J.A. Harvey, and A. Strominger,
``Quantum black holes,''\ajou Phys. Rev. &D46 (92) 638,
hep-th/9203042.}
\lref\Hawk{S.W. Hawking, ``Particle creation by black
holes,"\ajou Comm. Math. Phys. &43 (75) 199.}
\lref\Hawkii{S.W. Hawking, ``The unpredictability of quantum
gravity,''\ajou Comm. Math. Phys &87 (82) 395.}
\lref\BPS{T. Banks, M.E. Peskin, and L. Susskind, ``Difficulties for the
evolution of pure states into mixed states,''\ajou Nucl. Phys. &B244 (84)
125.}
\lref\Sred{M. Srednicki, ``Is purity eternal?,'' UCSB preprint
UCSBTH-92-22, hep-th/9206056.}
\lref\Cole{S. Coleman, ``Black holes as red herrings: Topological
fluctuations and the loss of quantum coherence,''\ajou Nucl. Phys. &B307
(88) 867.}
\lref\LoI{S.B. Giddings and A. Strominger, ``Loss of incoherence and
determination of coupling constants in quantum gravity,''\ajou Nucl. Phys.
&B307 (88) 854.}
\lref\CaWi{R.D. Carlitz and R.S. Willey, ``Reflections on moving
mirrors,''\ajou Phys. Rev. &D36 (87) 2327; ``Lifetime of a black
hole,''\ajou Phys. Rev. &D36 (87) 2336.}
\lref\Pres{J. Preskill, ``Do black holes destroy information?'' Caltech
preprint CALT-68-1819, hep-th/9209058.}
\lref\ACN{Y. Aharonov, A. Casher, and S. Nussinov, ``The unitarity
puzzle and Planck mass stable particles,"\ajou Phys. Lett. &B191 (87)
51.}
\lref\BDDO{T. Banks, A. Dabholkar, M.R. Douglas, and M O'Loughlin, ``Are
horned particles the climax of Hawking evaporation?'' \ajou Phys. Rev.
&D45 (92) 3607.}
\lref\BOS{T. Banks, M. O'Loughlin, and A. Strominger, ``Black hole remnants
and the information puzzle,'' hep-th/9211030, {\sl Phys. Rev. D} to
appear.}
\lref\GiMa{G.W. Gibbons and K. Maeda, ``Black holes and membranes in
higher-dimensional theories with dilaton fields,''\ajou Nucl. Phys. &B298
(88) 741.}
\lref\GHS{D. Garfinkle, G. Horowitz, and A. Strominger, ``Charged black holes
in string theory,''\ajou Phys. Rev. &D43 (91) 3140, erratum\ajou Phys. Rev.
& D45 (92) 3888.}
\lref\MSW{G. Mandal, A Sengupta, and S. Wadia, ``Classical solutions of
two-dimensional
string theory,'' \ajou Mod. Phys. Lett. &A6 (91) 1685.}
\lref\Witt{E. Witten, ``On string theory and black holes,''\ajou Phys. Rev.
&D44 (91) 314.}
\lref\RST{J.G. Russo, L. Susskind, and L. Thorlacius, ``Black hole
evaporation in 1+1 dimensions,''\ajou Phys. Lett. &B292 (92) 13,
hep-th/9201074.}
\lref\Suth{L. Susskind and L. Thorlacius, ``Hawking radiation and
back-reaction,''\ajou Nucl. Phys & B382 (92) 123, hep-th/9203054.}
\lref\Hawkiii{S.W. Hawking, ``Evaporation of two dimensional black
holes,''\ajou Phys. Rev. Lett. & 69 (92) 406,
hep-th/9203052.}
\lref\RSTii{J.G. Russo, L. Susskind, and L. Thorlacius, ``The endpoint of
Hawking radiation,''\ajou Phys. Rev. &D46 (92) 3444, hep-th/9206070}
\lref\deAl{S.P. deAlwis, ``Quantization of a theory of 2d dilaton
gravity,''\ajou Phys. Lett. &B289 (92) 278, hep-th/9205069; ``Black
hole
physics from Liouville theory,''
Boulder preprint COLO-HEP-284, hep-th/9206020.}
\lref\PiSt{T. Piran and A. Strominger, ``Numerical analysis of black hole
evaporation,'' ITP preprint NSF-ITP-93-36, hep-th/9304148}
\lref\RuTs{J.G. Russo and A.A. Tseytlin, ``Scalar-tensor quantum gravity
in two dimensions,'' Stanford/Cambridge preprint
SU-ITP-92-2=DAMTP-1-1992.}
\lref\BiCa{A. Bilal and C. Callan, ``Liouville models of black hole
evaporation,'' Princeton preprint PUPT-1320, hep-th/9205089.}
\lref\tHoo{G. 't Hooft, ``The black hole interpretation of string
theory,''\ajou Nucl. Phys. &B335 (90) 138.}
\lref\Page{D. Page, ``Black hole information,'' Alberta-Thy-23-93, to
appear in the proceedings of the
5th Canadian Conference on General Relativity and Relativistic Astrophysics,
Waterloo, Ontario, 1993 May 13-15, hep-th/9305040.}
\lref\VeVe{E. Verlinde and H. Verlinde, ``A unitary S matrix and 2-d black
hole formation and evaporation,'' Princeton preprint PUPT-1380, hep-th
9302022.}
\lref\SVV{K. Schoutens, E. Verlinde and H. Verlinde, ``Quantum black hole
evaporation,'' Princeton preprint PUPT-1395.}
\lref\StTr{A. Strominger and S. Trivedi, ``Information consumption by
Reissner-Nordstrom black holes,'' ITP/Caltech preprint
NSF-ITP-93-15=CALT-68-1851, hep-th/9302080.}
\lref\RuTs{J.G. Russo and A.A. Tseytlin, ``Scalar-tensor quantum gravity
in two dimensions,''\ajou Nucl. Phys. & B382 (92) 259.}
\lref\HVer{H. Verlinde, lectures, this volume.}
\lref\PaSt{Y. Park and A. Strominger, ``Supersymmetry and positive energy in
classical and quantum
two-dimensional dilaton gravity,''\ajou Phys. Rev. &D47 (93) 1569.}
\lref\NePa{W. Nelson and Y. Park, ``N=2 supersymmetry in two-dimensional
dilaton gravity,'' UCSB preprint UCSBTH-93-10, hep-th/9304163.}
\lref\Noji{S. Nojiri, ``Dilatonic supergravity in two-dimensions and the
disappearance of quantum
black hole,''\ajou Mod. Phys. Lett. &A8 (93) 53, hep-th/9209118.}
\lref\MaSh{E. Martinec and S.L. Shatashvili, ``Black hole physics and
liouville theory,''\ajou Nucl. Phys. &B368 (92) 338.}
\lref\Das{S.R. Das, ``Matrix models and black holes,''\ajou Mod. Phys.
Lett. &A8 (93) 69, hep-th/9210107.}
\lref\DMW{A. Dhar, G. Mandal, and S.R. Wadia, ``Stringy quantum effects in
two-dimensional black hole,''\ajou Mod. Phys. Lett. &A7 (92) 3703;
``Wave propagation in stringy black hole,'' Tata preprint TIFR-TH-93-05,
hep-th/9304072.}
\lref\JeYo{A. Jevicki and T. Yoneya, ``A deformed matrix model and
the black hole background in two-dimensional string
theory,'' ITP preprint NSF-ITP-93-67=Komaba preprint
UT-KOMABA 93/10, hep-th/9305109.}
\lref\HaSt{J.A. Harvey and A. Strominger, ``Quantum aspects of black
holes,'' preprint EFI-92-41, hep-th/9209055, to appear in the proceedings
of the 1992 TASI Summer School in Boulder, Colorado.}
\lref\Erice{S.B. Giddings, ``Toy models for black hole evaporation,''
UCSBTH-92-36, hep-th/9209113, to appear in the proceedings of the
International Workshop of Theoretical Physics, 6th Session, June 1992,
Erice, Italy.}
\lref\Dyso{F. Dyson, Institute for Advanced Study preprint, 1976,
unpublished.}
\lref\Astro{A. Strominger, ``Fadeev-Popov ghosts and 1+1 dimensional black
hole evaporation,'' UCSB preprint UCSBTH-92-18, hep-th/9205028.}


\Title{\vbox{\baselineskip12pt\hbox{UCSBTH-93-16}
\hbox{hep-th/9306041}
}}
{\vbox{\centerline {Black Holes and Quantum Predictability}
}}

\centerline{{\ticp Steven B. Giddings}\footnote{$^\dagger$}
{Email addresses:
giddings@denali.physics.ucsb.edu, steve@voodoo.bitnet.}
}

\vskip.1in
\centerline{\sl Department of Physics}
\centerline{\sl University of California}
\centerline{\sl Santa Barbara, CA 93106-9530}

\bigskip
\centerline{\bf Abstract}
A brief review of the confrontation between black hole physics and
quantum-mechanical unitarity is presented.  Possibile reconciliations
are modifying the laws of physics to allow
fundamental loss of information, escape of
information during the Hawking process, or black hole remnants.  Each of
these faces serious objections.
A better understanding of the
problem and its possible solutions can be had by studying two-dimensional
models of dilaton gravity.  Recent developments in these investigations are
summarized.  (Linear superposition of talks presented at the {\sl 7th
Nishinomiya Yukawa Memorial Symposium} and at the {\sl
1992 YITP Workshop on Quantum Gravity}, November 1992.)

\Date{}

A question on which there has been much recent discussion is that of
whether black holes lead to breakdown of quantum-mechanical
predictability.  This possibility was raised with Hawking's discovery
\refs{\Hawk} that black holes radiate.  I will begin by summarizing the
basic arguments on this issue.
Suppose that we start with a matter distribution of mass $M$ in
a pure quantum state and that we arrange for it to gravitationally
collapse to form a black hole.  The black hole then emits Hawking
radiation and loses mass.  Eventually it approaches the Planck mass,
$M_{p\ell}$, after which we do not know what happens.  Let us make the
most economical assumption (we will turn to others momentarily):  the
black hole evaporates completely leaving behind nothing but Hawking
radiation, as pictured in fig.~1.
This radiation is approximately thermal, and according to
Hawking's calculation is described by a mixed state.  Thus a pure state
has evolved into a mixed state.  Although we know not the reason,
unitary quantum mechanics must have failed, and an amount of quantum
information $\sim {M^2/M^2_{p\ell}}$ has been lost.\foot{Recall that
the quantum information in a density matrix $\rho$
for an $n$-state system is $\ln n +$Tr$\rho\ln\rho = \ln n - S$, where $S$
is the entropy.}

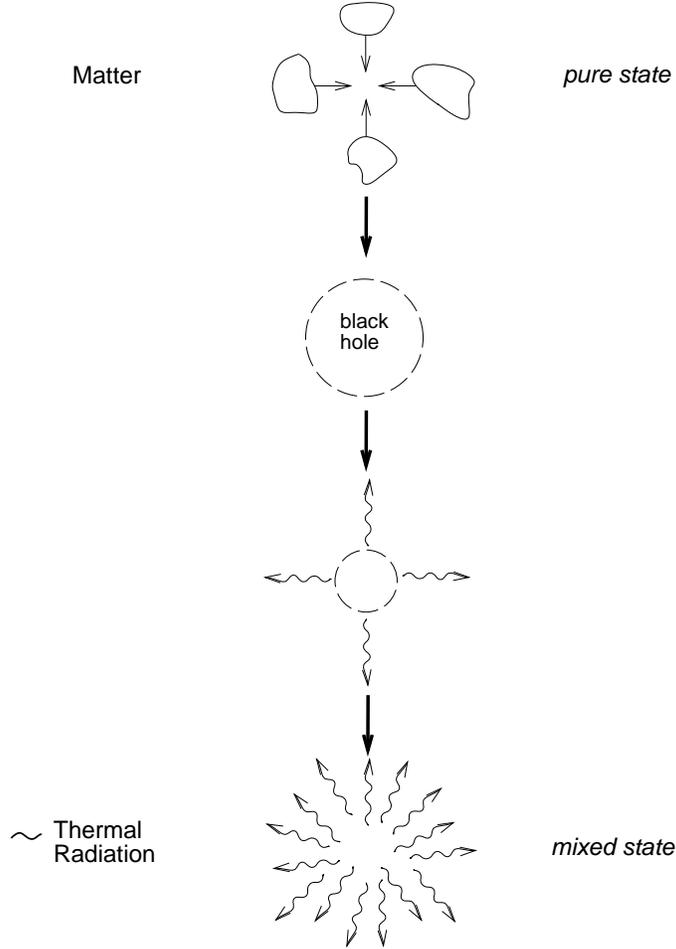
\ifig{\Fig\collaps}{A sketch of the collapse and evaporation process for
a black hole.  According to Hawking's original calculation, an initially
pure collapsing state of collapsing matter is converted into a mixed state
of nearly thermal Hawking radiation.}{japan.fig1}{5.00}

There are problems with this.  First, it is distasteful to violate a
central principle like unitarity.  And if we do allow such violations,
they might be expected to creep into other parts of physics besides
black holes.  Indeed, the general quantum principle that for any real
process there are similar virtual processes strongly suggests
that the
nonunitarity
should appear in ordinary physics at some level.  One might describe
this as resulting from black hole contributions to loop diagrams.
In fact Hawking proposed \refs{\Hawkii} that we revise the laws of
nature to incorporate such a breakdown of quantum predictability.
Specifically, he suggested that the quantum-mechanical evolution
operator $U=e^{-iHt}$ be replaced by an evolution operator $\Ssl$
acting linearly on density matrices,
\eqn\dollar{\rho\to \Ssl\rho\ .}
Such an evolution law can be arranged to conserve probability but
generically violates unitarity.

The objection to this was given by Banks, Peskin, and Susskind
\refs{\BPS} and by Srednicki \refs{\Sred}. One expects unitarity violation
to be ${\cal O}(1)$ at the Planck scale: once it is allowed
there is no obvious small
parameter by which it should be suppressed.  However, as
refs.~\refs{\BPS,\Sred} describe, evolution as in \dollar\  is only
nonunitary to the extent that it violates energy conservation:
information transfer requires energy transfer.  This implies a
disastrous breakdown of energy conservation resulting from incoherent
fluctuations of magnitude $\Delta E\sim M_{p\ell}$ in the energy;
roughly one such fluctuation occurs per Planck volume per Planck time.

We are thus forced to contemplate other alternatives; these are
summarized\foot{For other recent reviews of the information
problem see
\refs{\Pres\HaSt\Erice-\Page}.} in Fig.~2.
There are three logical possibilities:
information is either lost, re-emitted from the black hole,
or retained in some form of
remnant. We will examine these possibilities.

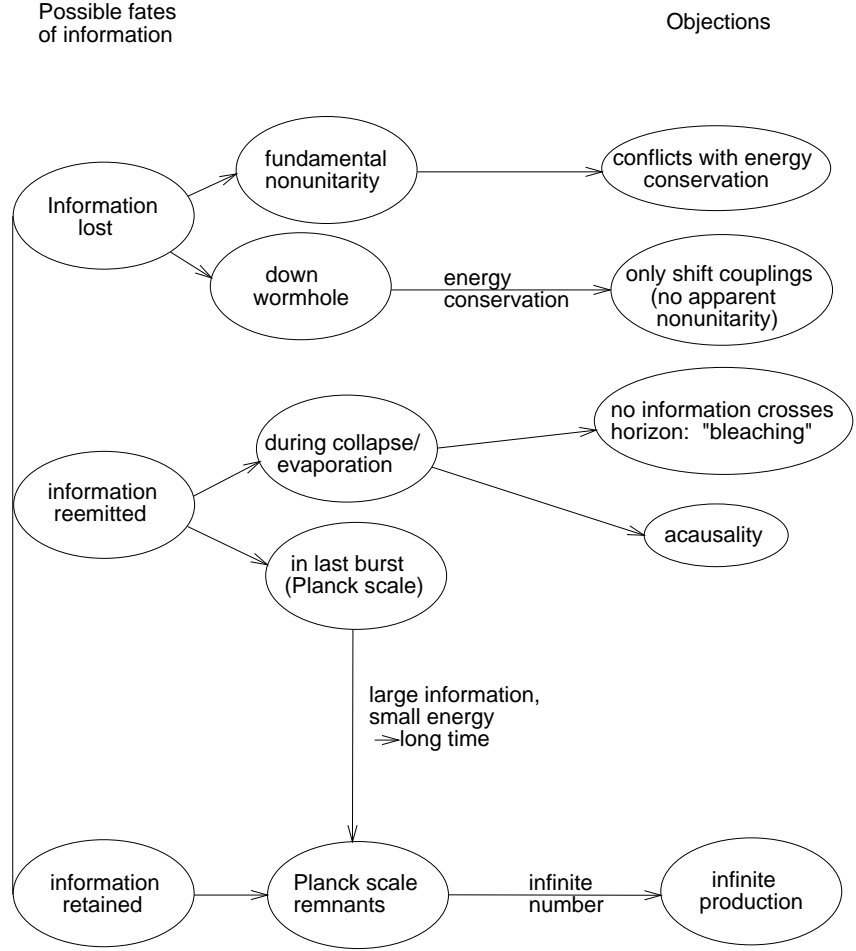
\ifig{\Fig\options}{The three logical possibilities for the fate of
information are shown on the left; there consequences are shown on the
right.  Each leads to a serious objection phrased solely in terms of
low-energy physics.}{japan.fig2}{5.0} 

First, if information is lost this may not be due to nonunitarity in the
laws of physics. It could simply be somewhere we can't find it. For
example, it could have been transmitted to a separate universe via a
spacetime wormhole\refs{\Dyso}.
However, this effective loss would be expected to
be described from our viewpoint by eq.~\dollar, and the conclusions
remain: effective loss of unitarity implies huge fluctuations in energy.
Conversely, if energy is conserved there is no apparent loss of
unitarity; indeed, wormholes just shift the constants of nature
\refs{\Cole,\LoI}. This is of no apparent use in explaining information
loss via black holes.

If information is re-emitted that can happen either during the Hawking
process or at the final burst when the black hole reaches the Planck
scale. In the first case this would require either that the information
never crosses the horizon or that it escapes from  inside the horizon.  For it
to not cross the horizon, all information must be ``bleached'' from the
infalling matter at the horizon.\foot{One might have alternatively hoped
that information about the infalling state is imprinted on the Hawking
radiation without anything special happening to the infalling state, \eg\
through an evolution law $|A\rangle_{\rm in}\rightarrow |A\rangle_{\rm out}
\otimes |A\rangle_{\rm black hole}$.  This however contradicts
superposition: only $|A\rangle_{\rm in}\rightarrow |A\rangle_{\rm out}
\otimes |I\rangle_{\rm black hole} $ is allowed for some state
$|I\rangle_{\rm black hole}$ independent of the initial state
$|A\rangle_{\rm in}$.  Put differently:  there are no quantum xerox
machines. }
However, for a large enough initial mass the horizon occurs at
arbitrarily weak curvature, and analyses of physics seen by infalling
observers show no special behavior there, making this alternative seem
implausible. If the information is to escape from inside
 before the final burst, it
must propagate acausally in regions of spacetime with weak curvature.
If such acausal propagation is allowed inside a black hole it should
by locality  likewise be allowed outside, which is unacceptable.
Finally, the information could emerge at the Planck scale where such
notions of causality may fail.  However, the remaining energy in the
black hole is $\sim M_{p\ell}$, and an amount of information
${M^2/M^2_{p\ell}}$  must be emitted.  This is possible, for example,
by radiating many very soft photons, but by basic quantum
mechanics \refs{\CaWi, \Pres} must
take a time $\sim M^4/M^4_{p\ell}\, t_{p\ell}$. There must therefore be
arbitrarily long-lived remnants.

Such remnants would have mass $\sim M_{p\ell}$ and must have an infinite
spectrum to encode the information from an arbitrarily large initial
black hole.\foot{For various remnant proposals see \eg\
\refs{\ACN\CGHS\BDDO-\DXBH}.}
Since large information content is their primary
characteristic, they will be referred to as {\it informons}.  Although the
rate for producing a given informon state in ordinary processes is
incredibly small, it is not zero.  The inclusive rate for producing all
such objects should then be infinite due to the infinite spectrum.  All
physical processes would produce bursts of remnants.

We now appear to be left with no palatable alternative.  This is the
black hole information conundrum.

A loophole should therefore appear in the above arguments.
For example, it's possible that
virtual processes do not feed information loss into
low-energy physics.  Alternately, there may be other descriptions of
information loss distinct from \dollar\ and without fluctuating energy.
Another possibility is that some theories of informons avert infinite
production.  One recent proposal\refs{\BOS} suggests that this might
occur for certain remnants in string theory
\refs{\GiMa,\GHS,\CGHS\BDDO-\DXBH}.
Although the details of this proposal appear flawed, it does suggest
modifications of effective field theories for remnants that might avoid
infinite production \refs{\CBHR}.  From a different angle, the arguments
about Hawking radiation test application of the equivalence principle in
the extreme:  the redshift between the horizon, where the radiation
originates, and infinity, is infinite.\foot{For a related discussion see
\refs{\Jaco}.}  Perhaps one should contemplate
the radical step of allowing
violations of the equivalence principle that become important only at high
energies.

In attacking these problems, a better understanding of black hole
evaporation would be useful.  This requires treating the complicated
dynamics of the backreaction of the Hawking radiation on the metric,
and ultimately of quantum gravity.  This may be more easily done by
stripping the problem to its bones:  we consider a two-dimensional toy
model \refs{\CGHS}
\eqn\dilact{
S= \otp\int d^2\sigma\,\sqrt{-g}\ e^{-2\phi}\left[\left(R+4(\nabla\phi)^2 +
4\lambda^2\right) - \half \sum\limits^N_{i=1} (\nabla f_i)^2\right]\ .
}
Here $\phi$ is a scalar dilaton, $\lambda^2$ a
coupling similar to the cosmological constant,
and $f_i$ are $N$ minimally coupled matter fields.  As we will see, this
model has black holes and Hawking radiation, and therefore contains the
essence of the information riddle.  It also has the virtues of being
classically soluble and renormalizable, and furthermore gives the
low-energy effective theory for certain four-dimensional black holes.

First consider the vacuum solutions; choose units so that $\lambda=1$
and define $\sigma^\pm = \tau \pm \sigma$. With $f_i=0$ the general
solution is
\eqn\three{
\eqalign{
ds^2 & = -\frac{d\sigma^+ d\sigma^-}{1+Me^{\sigma^--\sigma^+}}\cr
\phi & = -\half \ell n \left( M+e^{\sigma^+-\sigma^-}\right)\cr
}}
where $M$ is the mass.  For $M=0$ we have the ground state:
\eqn\LDV{
\eqalign{
ds^2 & = -d\sigma^+ d\sigma^-\cr
\phi & = -\sigma\ .\cr
}}
This is the closest analogue to four-dimensional Minkowski space, and is
called the linear dilaton vacuum. For $M>0$ the solution is a black hole
\refs{\MSW, \Witt} with Penrose diagram shown in Fig.~3. The horizon
is at $\sigma^+-\sigma^-\to-\infty$, and for future reference notice that
$e^{2\phi}|_{\rm horizon}=\frac{1}{M}$. For $M<0$ it is a naked
singularity.
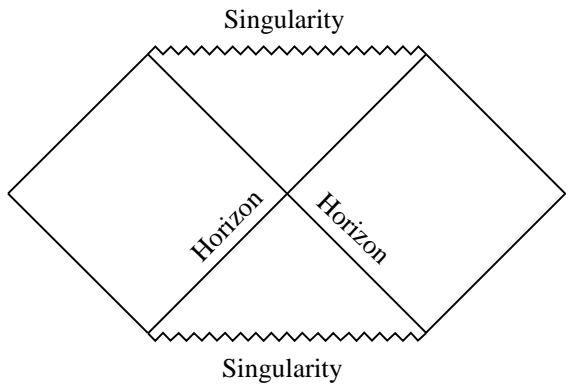
\ifig{\Fig\vpen}{Shown is the Penrose diagram for a vacuum
two-dimensional dilatonic black hole.}{japan.fig3}{2.00}

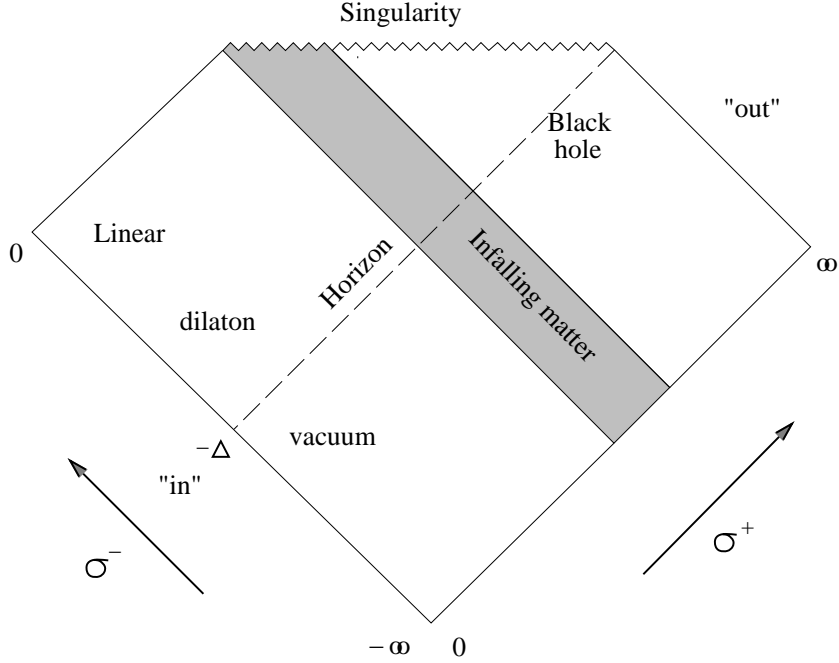
\ifig{\Fig\cpen}{The Penrose diagram for a collapsing black hole formed
from a left-moving matter distribution.}{japan.fig4}{3.50}

Next consider sending infalling matter, $f_i=F(x^+)$, into the \ldv.
This will form a black hole, as shown in Fig.~4.  Before the matter
infall the solution is given by \LDV. Afterwards it is
\eqn\fmet{\eqalign{ e^{-2\phi}&= M+
e^{\sigma^+}\left(e^{-\sigma^-}-\Delta\right)\cr
ds^2 &= -\frac{d\sigma^+ d\sigma^-}{1+M\,e^{\sigma^- - \sigma^+} - \Delta
\, e^{\sigma^-}}}
}
where
\eqn\six{
\eqalign{
M = & \int d\sigma^+ T_{++}\cr
\Delta = & \int d\sigma^+ e^{-\sigma^+}T_{++}\cr
}}
and
\eqn\seven{
T_{++}=\half(\partial_+ F)^2
}
is the stress tensor.  The coordinate transformation
\eqn\eight{
\xi^- = -\ell n\left(e^{-\sigma^-} - \Delta\right)
}
returns the metric to the asymptotically flat form
\eqn\nine{
ds^2 = -\frac{d\xi^+ d\xi^-}{1+M\ e^{\xi^- - \xi^+}}\ .
}

Hawking radiation can be described by observing that positive frequency
modes according to the future asymptotic observer,
\eqn\ten{
v_\omega = {1\over\sqrt{2\omega}} e^{-i\omega\xi^-}\ ,
}
are superpositions of both positive and negative frequency modes seen by
the observer in the \ldv.  The later are
\eqn\eleven{
\eqalign{
u_\omega = & {1\over\sqrt{2\omega}}e^{-i\omega\sigma^-}\cr
u_\omega^* = & {1\over\sqrt{2\omega}}e^{i\omega\sigma^-}\cr
}}
and the relation is
\eqn\twelve{
v_\omega = \int\nolimits^\infty_0 d\omega^\prime
\left(\alpha_{\omega\omega^\prime}
\ u_\omega^\prime + \beta_{\omega\omega^\prime}\ u^*_{\omega^\prime}\right)
}
for constants $\alpha_{\omega\omega^\prime}, \beta_{\omega\omega^\prime}$.
(Only the right-moving modes are relevant to the Hawking radiation.)
Therefore the in vacuum corresponds to an excited state in the out region;
indeed
\eqn\thirteen{
_{\rm in}\bigl\langle 0 |N^{\rm out}_\omega | 0\bigr\rangle_{\rm in}
= \int\nolimits^\infty_0 d\omega^\prime | \beta_{\omega\omega^\prime} |^2
}
for the out number operator $N^{\rm out}_\omega$.  The Bogoliubov
coefficients $\alpha_{\omega\omega^\prime}, \beta_{\omega\omega^\prime}$
can be found explicitly \refs{\GiNe}, and
\eqn\fourteen{
\beta_{\omega\omega^\prime} =
\frac{1}{2\pi}\,\sqrt{\frac{\omega^\prime}{\omega}}\ \Delta^{i\omega}
B\bigl(-i\omega-i\omega^\prime; 1 + i\omega\bigr)
}
where $B$ is the usual beta function.

The outgoing stress tensor for the Hawking radiation is likewise
computable.  For a general relationship between the initial and final
coordinates it is proportional to the schwarzian derivative,
\eqn\Schwarz{
_{\rm in}\bigl\langle 0 |T^f_{--} | 0\bigr\rangle_{\rm in}=
-{N\over24}\left\{-\hf \left[ \partial_{\xi_-} \ln\left({d\sigma^-\over
d\xi^-}\right)\right]^2 + \partial_{\xi_-}^2 \ln\left({d\sigma^-\over
d\xi^-}\right)\right\} \ .}
In our case this together with \eight\ gives
\eqn\hawkst{
_{\rm in}\bigl\langle 0 |T^f_{--} | 0\bigr\rangle_{\rm in} =
\frac{N}{48} \left[1-\frac{1}{(1+\Delta e^{\xi^-})^2}\right]\
}
which for $\xi^- \to \infty$ goes to the constant (thermal) value 1/48.

So far we have assumed that the geometry is fixed, independent of the
Hawking radiation.  To treat the backreaction of this radiation on the
metric it is most efficient to consider quantization via
the functional integral
\eqn\sixteen{
\int {\cal D}g {\cal D}\phi\, e^{iS_{\rm grav}[g, \phi]} \int {\cal
D}f\ e^{-\frac{i}{4\pi} \int d^2\sigma \sqrt{-g}\, \sum^N_{i=1} (\nabla
f_i)^2}}
where the action \dilact\ has been separated into the gravitational and
matter parts. The matter functional integral is well-studied; it can be
readily evaluated using properties of the trace anomaly.
This gives
\eqn\seventeen{
\int {\cal D}f\, e^{-\frac{i}{4\pi} \int d^2\sigma\ \sqrt{-g}
\ \sum\limits^N_{i=1} (\nabla f_i)^2} = e^{iNS_{PL}}
}
with
\eqn\eighteen{
S_{PL} = -\frac{1}{96\pi} \int\int \sqrt{-g}\ \ d^2\sigma
\ \sqrt{-g^\prime}\ \  d^2\sigma^\prime\ R(x)\ \sq^{-1} (x, x^\prime)
R(x^\prime)
}
the Polyakov-Liouville action; here $\sq^{-1}$ is the Green-function for
the d'Alem\-ber\-tian, $\sq$.

The effect of this action on the geometry is found by calculating its
stress tensor.  By a change of coordinates the metric can always be put
into conformal gauge, $g_{\mu\nu} = e^{2\rho} \eta_{\mu\nu}$, and in this
gauge
\eqn\nineteen{
\left\langle T^{\rm matter}_{--} \right\rangle = \frac{2\pi
N}{\sqrt{-g}}\ \frac{\delta S_{PL}}{\delta g^{--}} = \frac{N}{12}
\ \Bigl[\partial^2_- \rho - (\partial_- \rho)^2 - t_-(\sigma^-)\Bigr]
}
where $t_-$ is fixed by the initial boundary conditions.  For the collapsing
black hole, $t_-$ is zero by the condition of no Hawking
radiation in the \ldv, and \fmet\ gives
\eqn\twenty{
\left\langle T^{\rm matter}_{--} \right\rangle = \frac{N}{48}
\left[1-\frac{1}{(1+\Delta\, e^{\xi^-})^2}\right]
}
in agreement with \hawkst.  Therefore the Polyakov-Liouville action
precisely accounts for the Hawking flux.

The quantum theory with Hawking radiation and backreaction is thus
encoded in the functional integral
\eqn\gravint{
\int {\cal D}g\, {\cal D}\phi\ e^{iS_{\rm grav} + iNS_{PL}}\
}
or similar integrals giving correlation functions.
At present we are unable to fully quantize this system (more discussion
will follow shortly).  This motivates us to study it in a semiclassical
approximation.  Outside the horizon of a massive black hole,
\eqn\twentytwo
{e^{2\phi} < e^{2\phi}|_{\rm horizon} = \frac{1}{M} << 1\ .
}
Since $e^\phi$ plays the role of the gravitational coupling (that is
$e^{-2\phi}$ appears in the gravitational action, \dilact) the theory is
thus weakly coupled outside the horizon.  $S_{PL}$ is a subleading
contribution in $e^{2\phi}$, and there may be other contributions at the
same order, \eg\ from the measure in \gravint.  $S_{PL}$ can, however, be
made dominant by taking the number of matter fields large, $N>>1$,
simultaneously as $e^{2\phi}$ becomes small.  This justifies
approximating the functional integral \gravint\ by the solutions of the
semiclassical equations
\eqn\twentythree{
\eqalign{
0=& \frac{\delta}{\delta\phi}\ \left(S_{\rm grav} + NS_{PL}\right)\cr
0=& \frac{\delta}{\delta g^{\mu \nu}} \left(S_{\rm grav} +
NS_{PL}\right)\ .\cr
}}
In conformal gauge these equations are
\eqn\twentyfour{\eqalign{0=&-4\partial_+\partial_-\phi +
4\partial_+\phi\partial_-\phi +
2\partial_+\partial_-\rho +  e^{2\rho}\cr
0=T_{+-}& = e^{-2\phi}(2\partial_+\partial_-\phi - 4
\partial_+\phi\partial_-\phi - e^{2\rho})
- {N \over 12} \partial_+\partial_-\rho\cr
0=T_{++}& = e^{-2\phi}
(4\partial_+\phi\partial_+\rho -
2\partial^2_+\phi) + \half \partial_+ f\partial_+f\cr
                  & - {N \over 12}\left(\partial_+\rho\partial_+\rho -
\partial^2_+\rho + t_+(\sigma^+)\right)\ \cr}}
and similarly for $T_{--}$.
These equations still have the \ldv\ as a solution: the modification
$S_{PL}$ is quadratic in $R$, and $R$ vanishes in the \ldv.  However,
they are no longer soluble and must be treated by general arguments,
numerical techniques, studying related soluble models, and other
trickery \refs{\BDDO,\RST\Suth\Hawkiii\BGHS\RSTii\deAl\Lowe-\PiSt}.  The basic
picture that emerges is the following.

The matter collapses until it reaches a critical coupling,
$e^{2\phi_{cr}}= \frac{12}{N}$.  Here a singularity in the semiclassical
equations develops (the kinetic operator degenerates) and the
semiclassical approximation breaks down.  This singularity is hidden
behind an apparent horizon.  The latter is defined by the curve where
$(\nabla e^{-\phi})^2=0$, motivated by the analogy between $e^{-\phi}$ and the
radius of the two-spheres in four-dimensional spherically symmetric
spacetimes.  The black hole
radiates Hawking radiation and the apparent horizon recedes until it
approaches the singularity; this occurs as the mass radiated approaches
the initial mass.  Once they touch it is impossible to specify the
future evolution without understanding the strong-coupling physics ---
we come into the shadow of the singularity.  There is an effective
horizon, defined as the last light ray that escapes without hitting
strong coupling; above this we cannot make predictions.

These features are exhibited in a Kruskal diagram in Fig.~5, and in the
Penrose diagram of Fig.~6.  Notice that we can't go beyond $\phi_{cr}$
even in the \ldv: the physics is strongly coupled.  Therefore
spacetime to the left of $\phi_{cr}$ has been truncated
in these
diagrams.

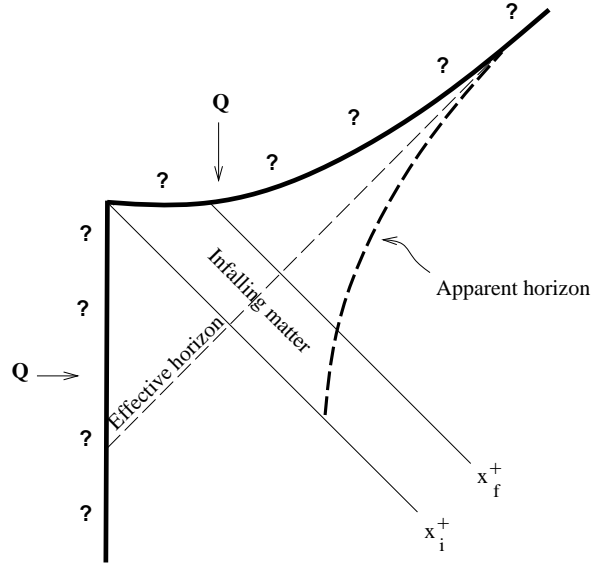
\ifig{\Fig\qkrus}{The Kruskal geometry of a collapsing and evaporating
black hole.  Past the line $Q$ the semiclassical approximation breaks down
and additional input is needed.}{japan.fig5}{3.00}

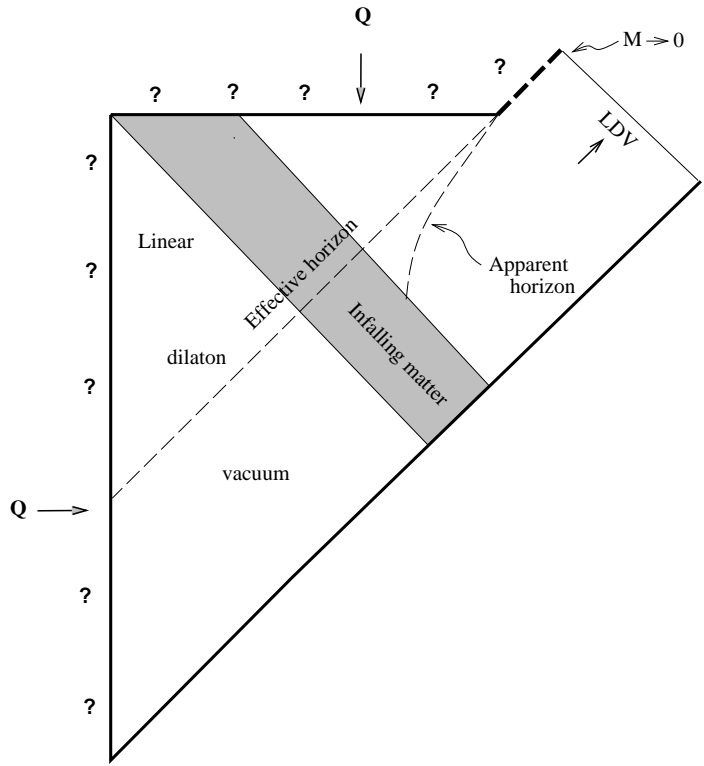
\ifig{\Fig\qpen}{The Penrose diagram corresponding to the geometry of
Fig.~5.  By the time the evaporation depends on strong-coupled physics at
the center of the black hole, most of the mass has been
radiated.}{japan.fig6}{4.00}

The semiclassical breakdown means that we cannot say what happens to the
information.  However, the explicit description of the evaporation,
which could be extended in a higher order analysis, gives one confidence
that the information does not emerge during the ordinary Hawking
evaporation.\foot{Although this appears to be the emerging consensus,
there are significant
holdouts; see \eg\ \refs{\tHoo\VeVe-\SVV}.}
This essentially follows from causality
together with the fact that the Hawking radiation emerges at weak
coupling.  Even more solid arguments for failure of information return have
been recently
constructed in similar two-dimensional models describing the s-wave
sector for Reissner-N\"ordstrom black holes\refs{\StTr}.
The picture is still, however, consistent with fundamental
information loss or long-lived remnants.

To do better would require a quantum treatment beyond the semiclassical
approximation.  I will briefly describe some of the issues in such a
treatment.

Quantization requires gauge fixing.  The most common approach is to fix
diffeomorphism invariance by picking a fixed background metric $\hat
g_{\mu\nu}$, and by restricting attention to metrics of the form
\eqn\gfix{
g_{\mu\nu} = e^{2\rho} \hat g_{\mu\nu}\ .
}
Thus the only ``dynamics'' of gravity is in the conformal factor $\rho$.
As at the outset, the action written in terms of $X^P = (\rho, \phi)$ is
renormalizable, but this isn't as useful in two dimensions as in four.
Although we shouldn't have to write down higher-dimension counterterms,
the fields $\rho$ and $\phi$ are dimension zero so we nonetheless expect
the full action to be of the form
\eqn\twentysix{
S = -\otp\int d^2\sigma\ \sqrt{-\hat g}\ \left[G_{MN}(X^P)\,\nabla
X^M\ \widehat\cdot\ \nabla X^N + \half \Phi (X^P) \hat R + T(X^P)\right]
}
where hat denotes quantities formed from the metric $\hat g$ and
$G_{MN}$, $\Phi$, and $T$ are general functions.  (The notation is
motivated by comparison to string theory.) An infinite number of
parameters are required to specify these functions, and thus the full
quantum theory.  Therefore although {\it renormalizable}, the theory is
uncomfortably close to four-dimensional gravity in being {\it
unpredictable}.

The functions $G_{MN}$, $\Phi$, and $T$ are not completely arbitrary,
but must satisfy certain constraints\refs{\QTDG}.  These are:

1. {\sl Background independence}. According to \gfix, the theory
should be left unchanged by the transformation
\eqn\twentyseven{
\eqalign{
\hat g_{\mu\nu} & \to e^{2\omega} g_{\mu\nu}\cr
 \rho          & \to \rho-\omega\ .\cr
}}
This condition (which is related to conformal invariance) implies
\eqn\twentyeight{
\eqalign{
\nabla_M \Phi \nabla^M T& -4T - \frac{\hbar}{2} \sq\, T +\cdots = 0\cr
\nabla_M\nabla_N \Phi& + \frac{\hbar}{2} R_{MN} + \cdots = 0\cr
(\nabla\Phi)^2 & - \frac{\hbar}{2}\, \sq\, \Phi + (N-24) \frac{\hbar}{3} +
\cdots = 0\cr
}}
where geometrical quantities are defined in terms of the metric $G_{MN}$,
$\hbar$ has been reinstated, and terms higher order in $\hbar$
have been dropped.  There are
infinitely many parameters determining solutions to these equations;
initial data for them could, for example, be specified by giving
$G_{MN}$, $\Phi$, and $T$ as functions of $\phi$ at a fixed scale,
$\rho$.

{2.} {\sl Classical Limit}.  As $e^\phi \to 0$, the theory should
agree with the classical theory, to leading order in $e^{2\phi}$:
\eqn\classS{
\eqalign{
G_{MN}& \to \left(\matrix{-4\,e^{-2\phi} & 2\, e^{-2\phi}\cr
                         2\, e^{-2\phi} & 0\cr}\right)\cr
\Phi & \to -2\, e^{-2\phi}\cr
T & \to -4 \lambda^2 e^{-2\phi}\cr
}}

{3.} {\sl Coupling to Hawking Radiation}. Hawking radiation should
cause black holes to shrink at the semiclassical rate, proportional to
the number $N$ of matter fields\refs{\Astro}.  Ghosts, which also contribute to
$S_{PL}$, should not affect the radiation rate of the black hole.

If the action is written as an expansion in $\hbar$ or equivalently
$e^{2\phi}$,
\eqn\thirty{
S= S_0 + S_1 +\cdots\ ,
}
with $S_0$ given by \classS, then this becomes a constraint on $S_1$.
With Polyakov-Liouville term included the one-loop action becomes
\eqn\thirtyone{
\eqalign{
\tilde S_1 =- \frac{\hbar}{2\pi} \int d^2\sigma\ \sqrt{-g}&
\left[G^{(1)}_{MN} \nabla X^M\ \widehat\cdot\ \nabla X^N + \half \Phi^{(1)}
(X^P) \hat R + T^{(1)} (X^P) \right]\cr
& - \frac{N-24}{96\pi}\hbar \int d^2\sigma\ \ \sqrt{-g}
\ d^2\sigma^\prime \sqrt{-\hat g^\prime}\ \hat
R\ \sq^{-1} \hat R\ .\cr
}}
(Here the -24 arises from the ghost contribution to the conformal anomaly.)
We want
\eqn\thirtytwo{
T_{--} = \frac{2\pi}{\sqrt{-\hat g}}\ \frac{\delta\tilde S_1}{\delta \hat
g^{--}}
}
to give the correct Hawking flux,$\rightarrow \hbar N/48$,
at infinity.  This requires
\eqn\thirtythree{
G^{(1)}_{\phi\phi} - \half \partial_\phi^2 \Phi^{(1)} = 2\ .
}
Finally there is

{4.} {\sl Stability}. The theory should have a stable ground state.

These conditions still leave an infinite family of theories, so one
requires still more input to fix a theory.  Several approaches have been
taken, none yet entirely satisfying.  In outline, they are

{1)} {\sl Quantum Soluble models}
\refs{\RuTs,\deAl,\BiCa,\RST,\QTDG,\VeVe,\SVV}
The classical metric $G_{MN}$ in \classS\ is flat.  For a variety of
solutions $G^{(1)}_{MN}$ can be seen not to spoil this.  Therefore, one
may choose coordinates $U(\rho, \phi)$, $V(\rho, \phi)$ so that $G_{UU}
= G_{VV}=0$ and $G_{UV}=\half$. It then turns out that the potential
term is also simple to leading order in $\hbar$; the action becomes
(here $\hat g_{\mu\nu} = \eta_{\mu\nu}$)
\eqn\thirtyfour{
S= \otp\int d^2\sigma\left(\nabla U\
\widehat\cdot\ \nabla V + \lambda^2 e^{2V}\right)
}
where non-trivial ${\cal O}(\hbar)$ corrections now appear in the
potential.  This theory has several virtues.  First, it satisfies 1--3
above. Secondly, this quantum action is of the same form as the
classical action \dilact\ in conformal gauge (with $V\leftrightarrow
\rho - \phi$ and $U\leftrightarrow e^{-2\phi}$) and thus is soluble.  Indeed,
the theory is a conformal field theory similar to the Liouville theory.
It furthermore has the property that the potential is not renormalized
since the propagator connects only $U$ and $V$ and the vertex contains
only $V$.  The major drawback is that Hawking radiation does not stop in
this model \refs{\deAl, \BiCa}, and indeed there are regular static
solutions with mass unbounded from below \refs{\QTDG}. Thus it fails
criterion 4.  Attempts have been made to remedy this by adding a
boundary condition that stabilizes the theory \refs{\RSTii,\HVer,\VeVe-\SVV}
with possibly interesting consequences.

{2)} {\sl Extra symmetry}. In the classical theory $j^\mu =
\partial^\mu(\rho-\phi)$ is conserved and one can try to preserve this
at the quantum level \refs{\RSTii,\HVer}. This does not, however, fix all
counterterms, and can run into the stability problem as above.
Alternately one can try supersymmetrization. The $N=1$ and $N=2$
theories have been considered \refs{\PaSt\Noji-\NePa} and there seem to be
applicable nonrenormalization theorems, but these appear still
insufficient to fully fix the theory \refs{\NePa}.

{3)} {\sl Strings}. We believe that string theory solves similar
problems in four-dimensional quantum gravity, so it is appealing to
apply it here.  Indeed, the low-energy string lagrangian in
two dimensions is
\eqn\thirtyfive{
S = \otp\int d^2 x \sqrt{-g}\ e^{-2\phi} \left[R+ 4(\nabla\phi)^2 +
4\lambda^2 - (\nabla T)^2 + T^2 + {\cal O}(T^{3}) +\cdots\right]\ ,
}
or, redefining the tachyon $T=e^\phi t$,
\eqn\stact{
\eqalign{
S&= \otp\int d^2 x\ \sqrt{-g}
\ \Bigl[e^{-2\phi}\left(R+ 4(\nabla\phi)^2 +
4\lambda^2\right) - (\nabla t)^2\cr
& + t^2 \left(\sq \phi - (\nabla\phi)^2
+ 1\right) + {\cal O} (e^\phi t^3) + \cdots\Bigr]\ .\cr
}}
The third term vanishes in the linear dilaton vacuum so $t$ behaves like
a massless scalar field, and the action \stact\ is strikingly similar to
\dilact.  Furthermore, matrix models are believed to provide a
consistent and essentially complete quantum description of
two-dimensional strings.  We might hope to use this to enlighten us on
black hole puzzles.

Unfortunately life is not so easy.  First, recall that the action
\stact\ is
valid only in the approximations
\item{i)} $e^\phi << 1$ $\ldots$ weak string coupling
\item{ii)} $k^\mu << 1$ $\ldots$ low momenta
\item{iii)} $T=e^\phi t << 1$ $\ldots$ weak tachyon;

\noindent
otherwise subleading terms become important.  We would like to know if
an object resembling a black hole, at least to the extent that it has an
apparent horizon, can be formed within this domain of validity.  To
examine this, use the fact that \stact\ is similar to dilaton gravity,
with $t \to f$, plus an extra repulsive potential.  Let's see if the
analogous conditions can be satisfied in dilaton gravity.  First, recall
$e^\phi < e^{\phi_h} = \frac{1}{\sqrt{M}}$, so condition\ i) is
satisfied for large black holes.  Next, we can arrange for \ ii) to be
satisfied by building the black hole from a large number of soft
particles.  But \ iii) is problematic: from \fmet-\seven\
the value of the dilaton at the
horizon is
\eqn\thirtyseven{
e^{-2\phi_h} \sim \int\nolimits^{\sigma^+_f}_{\sigma^+_i} d\sigma^+
e^{\sigma^+_i - \sigma^+} (\partial_+f)^2\ ,
}
where $\sigma_i^+$, $\sigma_f^+$ correspond to the beginning and end of the
pulse,
and this indicates that
\eqn\thirtyeight{e^{\phi_h} f\sim 1
}
at the horizon.  The same statement is likely true for the string
lagrangian; before a horizon forms and things get interesting, the
tachyon potential becomes important.\foot{I thank E. Martinec and A.
Strominger for discussions on this point.}

Since matrix models are supposed to include all higher order effects,
including the tachyon potential, one might hope to see evidence for or
against black holes in the matrix models.  However, at least naively
matrix models describe perturbations about the Liouville background
\eqn\thirtynine{
\eqalign{
T & = \mu e^{-\sigma} +\calo(e^{-2\sigma})\cr
\phi & = - \sigma + \frac{\mu^2}{8}\ e^{-2\sigma} +\calo(e^{-4\sigma}) \cr
g_{\mu v}& = \eta_{\mu\nu} +\calo(e^{-4\sigma})\cr
}}
with $\mu >> 1$. This means that when we consider such fluctuations
\eqn\forty{
T= \mu e^{-\sigma} + e^\phi t\ ,
}
the higher terms in the tachyon potential cause strong interactions
between the background and fluctuations.  For example, a $T^3$ term
gives a contribution
\eqn\fortyone{
\sim \mu e^{-\phi} (e^\phi t)^2\ .
}
Since $e^\phi t$ must become ${\cal O}(1)$ at the horizon, and since $\mu
>> 1$, this gets large {\it long} before the horizon forms.  The
Liouville wall obscures the black hole physics.  Therefore, it's quite
possible the matrix models simply describe reflection from the wall, not
black hole formation.  An optimist might hope for a non-perturbative
description of black holes to appear at $\mu \ltwid 1$, but this hope
has not yet been realized; a pessimist might worry that there is no
useful description of black hole formation in this
theory.\foot{As described in \refs{\MaSh\Das\DMW-
\JeYo}, an alternate interpretation of matrix models is in terms of
perturbations on a black hole background.  The connection of this to
black hole formation and to the above statements is unknown.}  It is
therefore not yet clear what strings might tell us about the problem of
formation and evaporation of black holes in two (or higher!) dimensions.

{}From this discussion we see that in some respects two-dimensional
dilaton gravity is perhaps more akin to four-dimensional gravity than we
might have hoped --- it still suffers unpredictability.  Nonetheless, we
have clearly made progress. We have a concrete and well-understood
semiclassical model for black hole evaporation embedded in a rich family
of theories.  Many of the complications of higher dimensions have been
stripped away, so we might hope to get to the essence of various
conceptual problems that arise in quantum gravity.  At the top of our
list is, of course, the issue of whether black holes avoid destroying
quantum information, and if so how they succeed.

\bigskip\bigskip\centerline{{\bf Acknowledgments}}\nobreak
I wish to thank my collaborators B. Birnir
C. Callan, J. Harvey, W. Nelson, and A. Strominger for the many insights
they have shared in the course of our work.
I also wish to thank T. Banks, E. Martinec, L. Susskind, L.
Thorlacius, S. Trivedi, and E. and H. Verlinde for
valuable discussions.
This work was supported in part by DOE grant DOE-91ER40618 and
by NSF PYI grant PHY-9157463.

\listrefs

\end